\documentclass[twocolumn]{aastex62}
\usepackage{xcolor}
\usepackage{apjfonts}

\newcommand{\cacnch}{Ca--CN--CH}
\newcommand{\cactio}{$Ca_{\rm CTIO}$}

\newcommand{\cajwl}{$Ca_{\rm JWL}$}
\newcommand{\chjwl}{$ch_{\rm JWL}$}
\newcommand{\cnjwl}{$cn_{\rm JWL}$}

\newcommand{\cnw}{CN-w}
\newcommand{\cni}{CN-i}
\newcommand{\cns}{CN-s}
\newcommand{\caw}{Ca-w}
\newcommand{\cas}{Ca-s}

\newcommand{\ebv}{$E(B-V)$}
\newcommand{\ecn}{$E$(\cnjwl)}
\newcommand{\ech}{$E$(\chjwl)}

\newcommand{\hst}{{\it HST}}
\newcommand{\kms}{km s$^{-1}$}

\newcommand{\nrgb}{$n$(G1):$n$(G2)}
\newcommand{\nrgbtwo}{$n$(\cnw):$n$(\cns)}
\newcommand{\nrgbthree}{$n$(\cnw):$n$(\cni):$n$(\cns)}

\newcommand{\pchjwl}{$\parallel$$ch_{\rm JWL}$}
\newcommand{\pcnjwl}{$\parallel$$cn_{\rm JWL}$}

\newcommand{\str}{Str\"omgren}

\newcommand{\vvhbmag}{$V - V_{\rm HB}$ $\leq$ 2.5 mag}
\newcommand{\cnwave}{$\lambda$3883}
\newcommand{\chwave}{$\lambda$4250}

\newcommand{\dy}{$\Delta Y$}

\newcommand{\hkjwl}{$hk_{\rm JWL}$}

\newcommand{\gaia}{{\it Gaia}}

\newcommand{\ocen}{$\omega$ Cen}

\shorttitle{M22}
\shortauthors{Lee}
\begin{document}

\title{Five Stellar Populations in M22 (NGC 6656)}

\author[0000-0002-2122-3030]{Jae-Woo Lee}
\affiliation{Department of Physics and Astronomy, Sejong University\\
209 Neungdong-ro, Gwangjin-Gu, Seoul, 05006, Korea\\
jaewoolee@sejong.ac.kr, jaewoolee@sejong.edu}

\begin{abstract}
We present the \cacnch\ photometry of the metal-complex globular cluster (GC) M22 (NGC 6656). Our photometry clearly shows the discrete double CN--CH anticorrelations in M22 red giant branch (RGB) stars, due to the difference in the mean metallicity. The populational number ratio between the two main groups is \nrgb\ = 63:37($\pm$3), with the G1 being more metal-poor. Furthermore, the G1 can be divided into two subpopulations with the number ratio of \nrgbtwo\ = 51:49 ($\pm$4), while the G2 can be divided into three subpopulations with \nrgbthree\ = 24:32:44 ($\pm$5). The proper motion of individual stars in the cluster shows an evidence of internal rotation, showing the G2 with a faster rotation, confirming our previous results from radial velocities. The cumulative radial distributions (CRDs) of individual subpopulations are intriguing in the following aspects: (1) In both main groups, the CRDs of the \cns\ subpopulations are more centrally concentrated than other subpopulations. (2) The CRDs of the the G1 \cns\ and the G2 \cns\ are very similar. (3) Likewise, the G1 \cnw\ and the G2 \cnw\ and \cni\ have almost identical CRDs. We also estimate the relative helium abundance of individual subpopulations by comparing their RGB bump magnitudes, finding that no helium abundance variation can be seen in the G1, while significant helium enhancements by \dy\ $\approx$ 0.03 -- 0.07 are required in the G2. Our results support the idea that M22 formed via a merger of two GCs.

\end{abstract}


\section{Introduction}
M22 (NGC 6656) is a metal-complex globular cluster (GC) in our Galaxy and its chemical peculiarity has been known for more than four decades. \citet{hesser79} noticed that M22 appears to have anomalies in its elemental abundances similar to \ocen. Later, \citet{norris83} reported the variation in the calcium abundance by up to $\Delta$[Ca/Fe] $\approx$ 0.3 dex, which was confirmed later by \citet{marino09,marino11} and \citet{jwlnat}.

The previous high-resolution spectroscopic studies clearly showed that M22 has a discrete bimodal metallicity distribution and it is an exemplar metal-complex GC\footnote{\citet{mucciarelli15} argued that RGB stars in M22 do not show any metallicity spread by reanalyzing the data presented by \citet{marino11}. As we showed in our previous work \citep{lee16}, several independent results from not only high- and low-resolution spectroscopy but also narrow and intermediate band photometry show strong evidence of metallicity spread in M22.} \citep{marino09,marino11,lee16}. In addition, \citet{lim15} reported the double CN--CH anticorrelations in M22 red giant branch (RGB) stars, which was lucidly interpreted by \citet{lee15} that they are natural consequences of the bimodal metallicity distribution.

From a photometric perspective, two or three groups of stars are classified in M22: \citet{marino09} identified the double sub-giant branch using the \hst\ F606W/F814W  photometry of the cluster. However, they did not list the populational number ratio. \citet{jwlnat} and \citet{lee15} employed the $hk$ photometry and they showed the discrete double RGB sequences due to the bimodal metallicity distribution. Later, \citet{milone17} showed that M22 contains at least three different groups of stars based on the so-called chromosome map (see their Figure 6).

During the past decade, we developed a new set of narrowband photometric systems in order to investigate multiple populations (MPs) in GC RGB and asymptotic giant branch (AGB) stars with small aperture telescopes \citep{lee15,lee17,lee18,lee19a,lee19b}. As we elaborately showed, our new photometric system allows us to measure accurate CN, CH, and calcium abundances even in the extremely crowded fields, such as the central part of GCs, where the traditional spectroscopic observations cannot be performed. It is a well-known fact that the nitrogen and carbon abundances can be altered through the CN cycle that occurred in the previous generation of stars, and the existence of the CN--CH anticorrelation indicates the presence of MPs in normal GCs. On the other hand, our photometric calcium abundance can tell the difference in metallicity among different populations. Consequently, our new photometric system is highly suitable for the study of MPs not only in normal GCs but also in metal-complex GCs.

In this Letter, we investigate the photometric CN--CH anticorrelations in M22, finding five MPs. As we will show later, our new discovery on the cumulative radial distributions (CRDs),  helium contents, and kinematical differences between individual populations strengthens our idea that M22 formed via a merger of two GCs \citep{lee15}.

\section{Observations}
The journal of observations of the $Ca_{\rm JWL}$ $by$ photometry is given in \citet{lee15}. In addition, we also obtained the $JWL39$ photometry using the CTIO 1 m telescope in three separate runs from 2013 April to 2014 May, and the $JWL43$ photometry using the KPNO 0.9 m telescope in two separate runs from May and September in 2018. The updated total integration times for our observations are given in Table~\ref{tab:obs}. 

The detailed discussion for our new filter system can be found in \citet{lee15,lee17,lee19a,lee19b}. The CTIO 1.0 m telescope was equipped with an STA 4k $\times$ 4k CCD camera, providing a plate scale of 0\farcs289 pixel$^{-1}$ and a field of view (FOV) of 20\arcmin\ $\times$ 20\arcmin. We obtained the photometry for the \str\ $uvby$, \cactio, \cajwl, and $JWL39$ using the CTIO 1.0m telescope with the mean airmass of 1.068 $\pm$ 0.072, and the combined FOV of our mosaicked science frames from CTIO runs was 1\arcdeg\ $\times$ 1\arcdeg. The KPNO 0.9 m telescope was equipped with the Half Degree Imager (HDI), providing a plate scale of 0\farcs43 pixel$^{-1}$  and a FOV of 30\arcmin\ $\times$ 30\arcmin, and we obtained \str\ $by$ and $JWL43$ using the KPNO 0.9m telescope. Since the altitude of M22 from KPNO is very low, with the maximum altitude of about 34\arcdeg, we paid special attention to acquire the correct extinction coefficients for each filter. The range of airmasses of the photometric standards for the \str\ $by$ and $JWL43$ filters was from 1.026 to 1.764 for the 2018 May run, and from 1.029 to 1.846 for the 2018 September run. For our M22 field, the range of airmass was from 1.780 to 1.819, similar to the maximum airmasses of the photometric standards. Also, due to the narrow bandwidth of our $JWL43$ filter, the color dependency of the extinction coefficient is negligibly small. Therefore, it is believed that our \chjwl\ measurements are correct.

The raw data handling was described in detail in our previous works \citep{lee15,lp16,lee17}. The photometry of M22 and standard stars were analyzed using DAOPHOTII, DAOGROW, ALLSTAR and ALLFRAME, and  COLLECT-CCDAVE-NEWTRIAL packages \citep{pbs87,pbs94,lc99}.

Finally, we derived the astrometric solutions for individual stars using the data extracted from the Naval  Observatory Merged Astrometric Dataset \citep{nomad} and the IRAF IMCOORS package.

\begin{deluxetable}{ccccc}
\tablenum{1}
\tablecaption{Integration times (s) for M22\label{tab:obs}}
\tablewidth{0pc}
\tablehead{
\multicolumn{5}{c}{New Filters} \\
\cline{1-5}
\colhead{$y$} &
\colhead{$b$} & 
\colhead{$Ca_{\rm JWL}$} &
\colhead{$JWL39$} &
\colhead{$JWL43$} 
}
\startdata
14705 & 32095 & 126740 & 25950 & 9650 \\
\enddata 
\end{deluxetable}

\begin{figure}
\epsscale{1}
\figurenum{1}
\plotone{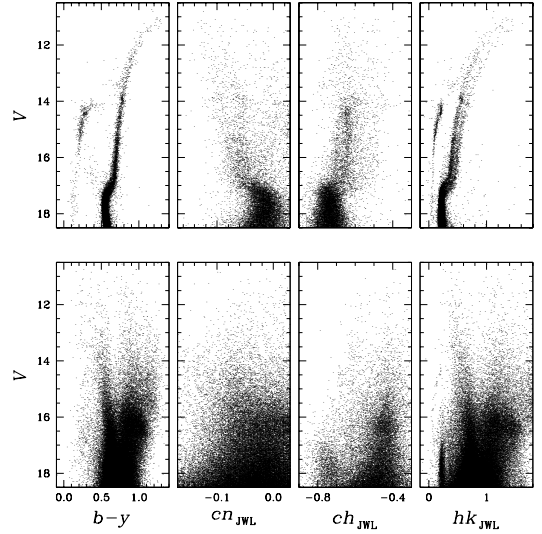}
\caption{
(Top panels) CMDs of M22 membership stars based on the proper motion study of the \gaia\ DR2. A weak bimodal RGB sequence can be seen in the M22 \cnjwl\ CMD, while a broad RGB sequence can be seen in the \chjwl\ CMD.
(Bottom panels) CMDs of the off-cluster field stars.
}\label{fig:cmd}
\end{figure}

\section{Results}
\subsection{Color--Magnitude Diagrams}
In Figure~\ref{fig:cmd}, we show color--magnitude diagrams (CMDs) of  bright stars in the M22 field \citep[see also][]{lee15}. Using the second \gaia\ date release \citep[\gaia\ DR2;][]{gaiadr2} and our multicolor photometry \citep[see, e.g.,][]{lee15}, we removed the off-cluster field stars and selected M22 membership RGB stars \citep[e.g., see][]{milone18,lee19b}.

Same as our previous work \citep{lee19b}, the definitions of photometric indices used in this work are
\begin{eqnarray}
cn_{\rm JWL} &=& JWL39 - Ca_{\rm JWL}, \label{eq:cn} \\
ch_{\rm JWL} &=& (JWL43 - b) - (b-y). \label{eq:ch}
\end{eqnarray}
The \cnjwl\ and \chjwl\ were introduced by the author of the paper and they are excellent photometric measures of the CN band at \cnwave\ and CH G band at \chwave, respectively, for cool stars \citep{lee17,lee18,lee19a,lee19b}.
We note that color excesses of our indices are relatively small, \ecn\ = 0.046$\times$\ebv\ and \ech\ = $-$0.418$\times$\ebv, calculated using the method described by \citet{lee01}, which make our indices less sensitive to variation in foreground reddening. For example, we estimated the degree of variation in foreground reddening of M22 by calculating the $(b-y)$ widths of RGB stars in each group (see below for the definition of the two RGB groups), obtaining $\sigma$\ebv\ $\approx$ 0.030 mag, which results in $\sigma$\ecn\ $\approx$ 0.001 mag and $\sigma$\ecn\ $\approx$ $-$0.013 mag, values too negligibly small to affect our results presented in this work.

The RGB sequences were parallelized using the following relation \citep[also see][]{milone17,lee19a,lee19b},
\begin{equation}
\parallel{\rm CI}(x) \equiv \frac{{\rm CI}(x) - {\rm CI}_{\rm red}}
{{\rm CI}_{\rm red}-{\rm CI}_{\rm blue}},\label{eq1}\label{eq:pl}
\end{equation}
where CI$(x)$ is the color index of individual stars and CI$_{\rm red}$, CI$_{\rm blue}$ are color indices for the fiducials of the red and blue sequences of individual color indices.

\begin{deluxetable}{lcccccc}
\tablenum{2}
\tablecaption{Populational number ratios (\%)\label{tab:pop}}
\tablewidth{0pc}
\tablehead{
\multicolumn{1}{c}{} &
\multicolumn{2}{c}{G1} &
\multicolumn{1}{c}{} &
\multicolumn{3}{c}{G2} \\
\cline{2-3}\cline{5-7}
\colhead{} &
\colhead{\cnw} &
\colhead{\cns} & 
\colhead{} &
\colhead{\cnw} &
\colhead{\cni} &
\colhead{\cns} 
}
\startdata
All     & 32.4 & 30.9 & & 8.7  & 12.0 & 16.0 \\
G1 only & 51.2 & 48.8 & & \nodata & \nodata & \nodata \\
G2 only & \nodata & \nodata & & 23.8 & 32.6 & 43.6 \\
\enddata 
\end{deluxetable}

\begin{deluxetable}{lcc}
\tablenum{3}
\tablecaption{$p$-values (\%) for Two Sample $t$-tests for G2 Subpopulations\label{tab:ttest}}
\tablewidth{0pc}
\tablehead{
\colhead{} &
\colhead{G2 \cni} &
\colhead{G2 \cns}
}
\startdata
G2 \cnw\ & 0.00 & 0.00 \\
G2 \cni\ &      & 0.35 \\
\enddata 
\end{deluxetable}

\begin{figure}
\epsscale{1}
\figurenum{2}
\plotone{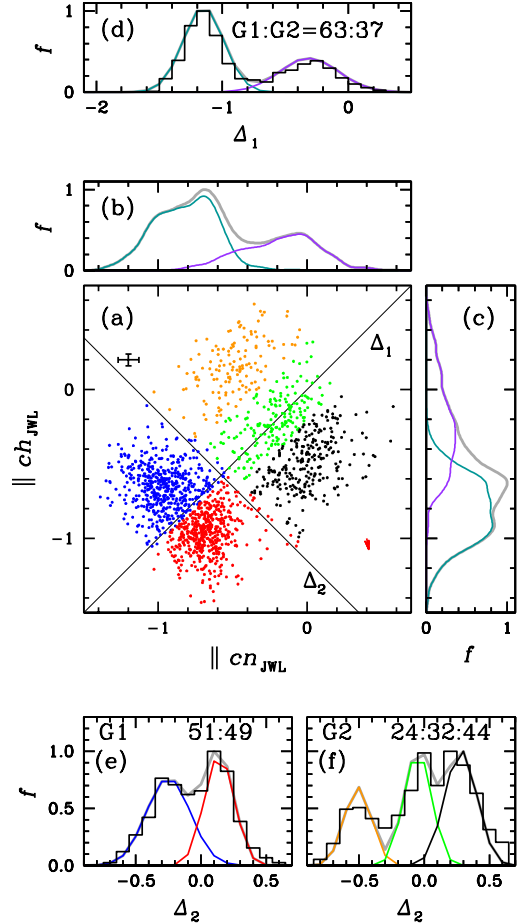}
\caption{
(a) Plot of the \pcnjwl\ versus \pchjwl\ of M22 RGB stars with \vvhbmag. The $\Delta_{\rm 1}$ indicates the axis where the decomposition of the main groups, the G1 and G2, is performed. On the other hand, the $\Delta_{\rm 2}$ indicates the axis where decompositions of the subpopulations of the G1 and G2 are performed. The mean measurement errors are also given  with black error bars. We also show the differential reddening vector with \ebv\ = 0.030 mag is shown with a red arrow, negligibly small to affect our results.
(b) The \pcnjwl\ distributions of the RGB stars. The dark green and purple colors indicate the G1 and G2 populations. Unlike the normal GCs without metallicity spread, such as M5 and NGC 6752, discrete separations in individual populations cannot be seen.
(c) Same as (b), but for the \pchjwl\ distribution.
(d) The $\Delta_{\rm 1}$ distribution. The dark green and purple colors are for the G1 and G2 populations, respectively. The discrete double RGB subpopulations between the G1 and G2 populations can be seen with the number ratio of \nrgb\ = 63:37.
(e) The $\Delta_{\rm 2}$ distribution of the G1 population. The two components can be seen with the number ratio of \nrgbtwo\ = 51:49.
(f) The $\Delta_{\rm 2}$ distribution of the G2 population. The Gaussian decomposition with three components can reasonably reproduce the observed distribution with the \nrgbthree\ = 24:32:44.
}\label{fig:cnch}
\end{figure}

\subsection{Populational Tagging from the \pcnjwl\ versus \pchjwl}
In our previous studies \citep[e.g., see][]{jwlnat,lee15}, we reported the bimodal calcium distribution of M22 RGB stars (namely, the \caw\ and \cas\ groups) based on their photometric calcium abundances in the $\Delta hk$ versus $V$ CMD (see the top rightmost panel of Figure~\ref{fig:cmd}), which is consistent with high-resolution spectroscopic studies showing the bimodal metallicity distribution of M22 \citep{marino09,marino11,lee16}.

In our current study, we perform populational tagging on the \pcnjwl\ versus \pchjwl\ plane. In Figure~\ref{fig:cnch}, we show the plot of the \pcnjwl\ versus \pchjwl\ of the M22 RGB stars with \vvhbmag. At first glance, two main groups of stars with their own photometric CN--CH anticorrelations can be seen, similar to what can be found in the low-resolution spectroscopic study of the cluster \citep{lim15}. As we already discussed in detail \citep{lee15}, the difference in the mean metallicity of the two main groups of stars is mainly responsible for these two separate CN--CH anticorrelations in M22: a group of RGB stars with low \pcnjwl\ and \pchjwl\ values is the lower-metallicity population (G1: the blue and red dots in Figure~\ref{fig:cnch} and the definition will be given below), while that with large \pcnjwl\ and \pchjwl\ corresponds to the higher-metallicity population (G2: the black, green, and orange dots). We emphasize that, due to the presence of the the multiple subpopulations in M22, and how they lie in this diagram, as well as photometric errors, clear populational separations in the \pcnjwl\ and \pchjwl\ distributions cannot be seen as shown in Figure~\ref{fig:cnch} (b-c).

In order to derive the two main groups of stars, we calculated the RGB distribution projected onto the $\Delta_{\rm 1}$ axis, which is a slope of 1, and we show our result in Figure~\ref{fig:cnch}(d). The $\Delta_{\rm 1}$ distribution of RGB stars shows a well-separated bimodal distribution. We employed the expectation maximization (EM) algorithm for the multiple-component Gaussian mixture distribution model to perform the populational tagging. We calculated the probability of individual RGB stars for being the G1 (i.e., RGB stars with smaller $\Delta_{\rm 1}$ values) and G2 (i.e., RGB stars with larger $\Delta_{\rm 1}$ values) groups in an iterative manner, where stars with $P($G1$|x_i)$ $\geq$ 0.5  from the EM estimator are denoted with the solid dark green lines, which corresponds to  the G1 population, while $P($G2$|x_i)$ $>$ 0.5 with the solid purple lines, which corresponds to the G2 population. Through this process, we obtained the RGB populational number ratio of \nrgb\ = 63:37 ($\pm$3), which is in excellent agreement with that by \citet{milone17}, who obtained $N_{\rm Type II}/N_{\rm TOT}$ = 0.403 $\pm$ 0.021, where our G2 corresponds to the Type II classified by \citet{milone17}.

It is worth noting the absence of any clear subpopulational \pcnjwl\ separations in the G1 (dark green) and G2 (purple) in M22, as shown in Figure~\ref{fig:cnch}(b), which is in sharp contrast to the normal GCs without metallicity spread (e.g.\ M3, M5, NGC 6723, and NGC 6752) exhibiting discrete double \cnjwl\ or \pcnjwl\ RGB sequences in our previous studies \citep[e.g., see][]{lee17,lee18,lee19a,lee19b}. Instead, the G1 and G2 distributions projected onto the $\Delta_{\rm 2}$ axis with the slope of $-$1 (i.e., on the line along the CN--CH anticorrelation) exhibit the double and triple peaks, respectively. Using multiple Gaussian decompositions, we obtained the subpopulational number ratios of \nrgbtwo\ = 51:49 ($\pm$4) for the G1 group and \nrgbthree\ = 24:32:44 ($\pm$5) for the G2 group, and we show our results in Figure~\ref{fig:cnch}(e)--(f). For G2 subpopulations, we performed Welch's two sample $t$-tests to see if they are drawn from the same population and we show $p$-values in Table~\ref{tab:ttest}, suggesting that they are different subpopulations. In the G1 group, the fraction of the \cnw, $\approx$ 0.50, is rather large compared to those of normal GCs with intermediate to high total masses, $\approx$ 0.30 \citep[e.g., see][]{lee17,lee18,lee19a,lee19b,milone17}. We note that the populational characteristic of the M22 G1 group (the subpopulational number ratio and the CRDs with a strong radial gradient as will be discussed below) is very similar to that of M3 \citep{lee19a}.

\subsection{Internal Rotation}
\begin{figure}
\epsscale{1}
\figurenum{3}
\plotone{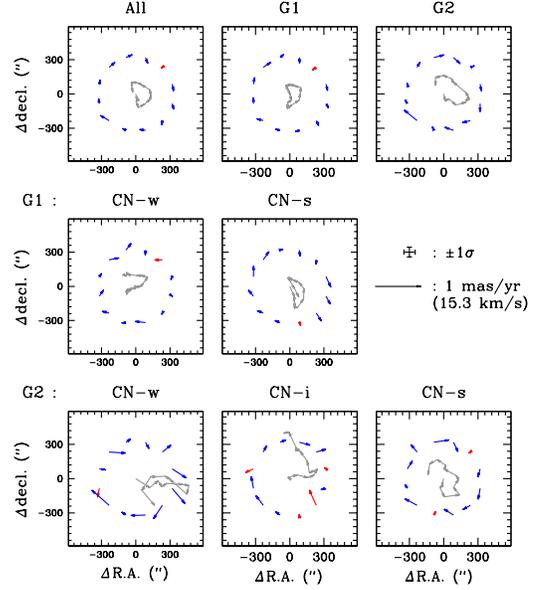}
\caption{
Distributions of the mean proper motions of 12 slices in the radial zone of 0\farcm5\ $\leq$ $r$ $<$ 10\arcmin. The red color denotes a clockwise rotation (E $\rightarrow$ N $\rightarrow$ W $\rightarrow$ S $\rightarrow$ E), while the blue color denotes a counterclockwise rotation at a given position vector. The gray arrows show evolutions of tangential vectors of consecutive slices in a counterclockwise sense starting at East.  The G2 appears to have larger projected tangential velocities than the G1 does.
}\label{fig:rot}
\end{figure}

We explore the internal rotation of individual subpopulations based on the proper motion study of the \gaia\ DR2 \citep{gaiadr2}. We divided the sphere into 12 different slices in a single radial zone of 0\farcm5\ $\leq$ $r$ $<$ 10\arcmin. Then, we calculated the mean proper motion vectors in each slice and show our results in Figure~\ref{fig:rot}. In the figure, we also show evolutions of tangential vectors of consecutive slices in a counterclockwise sense starting at east, where the size of the shape may indicate the degree of the internal rotation, although, for example, the G2 \cni\ does not show a closed loop. Our results show that M22 has a substantial internal rotation \citep[see also][]{sollima19}. 
We estimated rotational velocities of 1.9 $\pm$ 0.1 \kms\ and 2.4 $\pm$ 0.4 \kms for the G1 and G2, respectively, indicating that the G2 appears to have a slightly greater degree of internal rotation than the G1, consistent with our previous results from the radial velocity measurements \citep{lee15}.

\begin{figure}
\epsscale{1}
\figurenum{4}
\plotone{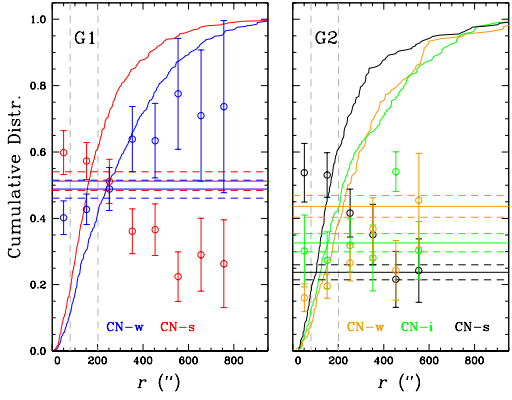}
\caption{
(Left) CRDs of the \cnw\ (blue) and \cns\ (red) in the G1 group. The G1 \cns\ is more centrally concentrated with a strong radial gradient. The vertical gray dashed lines denote the core and the half-light radii of the cluster. The horizontal solid lines denote the mean fractions of individual subpopulations, while the horizontal dashed lines the 1$\sigma$ error of the mean.
(Right) The CRDs of the \cnw\ (orange), \cni\ (green), and \cns\ (black) in the G2 group. Again, the G2 \cns\ is more centrally concentrated with a strong radial gradient.
}\label{fig:radial}
\end{figure}

\begin{deluxetable}{ccccc}[t]
\tablenum{4}
\tablecaption{ K-S Tests for Cumulative Radial Distributions\label{tab:ks}\tablenotemark{1}}
\tablewidth{0pc}
\tablehead{
\multicolumn{3}{c}{Populations} &
\multicolumn{1}{c}{$p$-value (\%)} &
\multicolumn{1}{c}{$D$}
}
\startdata
G1 & vs.\ & G2 & 29.2 & 0.052 \\
G2(\cnw) &  vs.\ & G2(\cni) & 32.6 & 0.101 \\
G2(\cnw) & vs.\ & G2(\cni\ + \cns) & 0.2 & 0.181 \\
G1(\cnw) & vs.\ & G2(\cnw) & 61.5 & 0.072 \\
G1(\cnw) & vs.\ & G2(\cni) & 59.0 & 0.066 \\
G1(\cns) & vs.\ & G2(\cns) & 62.4 & 0.058 \\
G1(\cnw) & vs.\ & G2(\cnw\ + \cni) & 85.7 & 0.043 \\
G1(\cns) & vs.\ & G2(\cni\ + \cns) & 4.2 & 0.091 \\
\enddata 
\tablenotetext{1}{Inter-subpupulational comparisons with the $p$-value of 0.0\% are omitted.}
\end{deluxetable}

\subsection{Cumulative Radial Distributions}
The CRDs of individual populations in GCs may provide a crucial information on the long-term dynamical evolution of GCs \citep[e.g., see][]{vesperini13}. For normal GCs without any perceptible metallicity spread, the CRDs of the \cnw\ and \cns\ populations in M5, NGC 6723, and NGC 6752 are very similar and statistical tests suggest that their \cnw\ and \cns\ populations are most likely drawn from same parent distributions \citep{lee17,lee18,lee19b}. On the other hand, the \cns\ population in M3 shows a more centrally concentrated CRD \citep{lee19a}.

Here, we derived the CRDs of individual subpopulations in M22 based on our photometric CN and CH abundances, and we obtained very intriguing results. 
First, the CRDs of the G1 and G2 groups are similar. We performed Kolmogorov--Smirnov (K-S) tests to derive the significance level for the null hypothesis that both distributions are drawn from the same distribution. We show the results for various cases in Table~\ref{tab:ks}. Our K-S tests show that the G1 and G2 are most likely drawn from the same parent distribution with a $p$-value of 29.2\%. 
Secondly, the \cns\ subpopulations in both the G1 and G2 groups are more centrally concentrated with a strong radial gradient  as shown in Figure~\ref{fig:radial}, which is a very distinctive feature of M3 compared to other normal GCs \citep{lee19a}.
Finally, the CRD of the G1 \cnw\ is very similar to those of the G2 \cnw\ and \cni, while the CRD of the G1 \cns\ is very similar to that of the G2 \cns.

\begin{figure}
\epsscale{1}
\figurenum{5}
\plotone{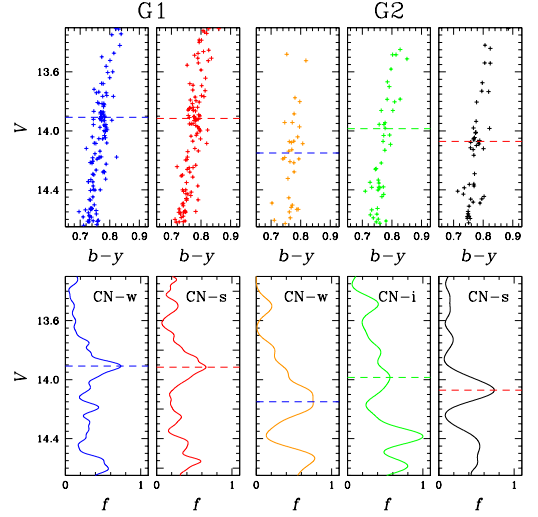}
\caption{
(Top panels) CMDs around the RGBB region. The horizontal dashed lines denote the RGBB $V$ magnitude.
(Bottom panels) Generalized differential luminosity functions. 
}\label{fig:rgbb}
\end{figure}

\begin{deluxetable}{clc}[t]
\tablenum{5}
\tablecaption{ RGB Bump magnitudes\label{tab:rgbb}}
\tablewidth{0pc}
\tablehead{
\multicolumn{2}{c}{Populations} &
\multicolumn{1}{c}{$V$} 
}
\startdata
G1 & \cnw & 13.908 ($\pm$0.025)\\
G1 & \cns & 13.915 ($\pm$0.025) \\
G2 & \cnw & 14.150 ($\pm$0.040)\\
G2 & \cni & 13.985 ($\pm$0.040)\\
G2 & \cns & 14.071 ($\pm$0.040)\\
\enddata 
\end{deluxetable}

\subsection{Red Giant Branch Bump Magnitudes}
During the evolution of the low-mass stars, the RGB stars experience slower evolution and temporary drop in luminosity when the very thin H-burning shell crosses the discontinuity in the chemical composition and lowered mean molecular weight left by the deepest penetration of the convective envelope during the ascent of the RGB, the so-called RGB bump \citep[RGBB; e.g., see][]{cassisi13}. The RGBB luminosity increases with helium abundance and decreases with metallicity at a given age.

We compared the RGBB $V$ magnitudes in order to understand the relative metallicity and helium abundance between individual subpopulations.
Figure~\ref{fig:rgbb} and Table~\ref{tab:rgbb}  show our results. The G1 \cnw\ and \cns\ have almost the same RGBB $V$ magnitudes,\footnote{Note that the G1 \cnw\ and \cns\ have very similar \hkjwl\ strengths at a given $V$ magnitude and, therefore, they have very similar metallicity. On the other hand, at a given $V$ magnitude, the G2 group has a larger \hkjwl\ value than the G1 and, therefore, the G2 is more metal rich \citep{lee15,lee16}.} a strong observational line of evidence that both subpopulations have the same metallicity and helium abundance, in sharp contrast to normal GCs, such as M5, NGC 6723, and NGC 6752, with discernible helium enhancements in their \cns\ populations \citep{lee17,lee18,lee19b}. Our result poses a strong constraint on the polluter of the chemical evolution of the G1 group: no helium enhancement but variations in C and N. Furthermore, the extent of the C and N variations in the G1 group is smaller than that in the G2 group.

The RGBB of the G2 \cnw\ is 0.242 $\pm$ 0.047 mag fainter than the the G1 group, which can be translated into the metallicity difference of $\Delta$[Fe/H] $\approx$ 0.26 $\pm$ 0.05 dex\footnote{In our previous work \citep{lee15}, we derived a relation between the RGBB magnitude versus metallicity using results by \citet{bjork06}, finding  $\Delta M_{V,{\rm bump}}/\Delta$[Fe/H] $\approx$ 0.93 mag/dex.} if there was no helium enhancement, in the sense that the G2 \cnw\ is more metal rich than the G1. 
Our photometric estimate of metallicity difference is slightly larger than that of \citet{marino11}, who obtained the metallicity difference between the two groups of stars in M22, $\Delta$[Fe/H] $\approx$ 0.15 $\pm$ 0.02 dex, by employing high-resolution spectroscopy. 
On the other hand, \citet{lee16} employed the line-by-line differential spectroscopic analysis, obtaining $\Delta$[Fe/H]$_{\rm I}$ = 0.20 $\pm$ 0.04 dex and $\Delta$[Fe/H]$_{\rm II}$ = 0.17 $\pm$ 0.06 dex, in good agreement with that from RGBB $V$ magnitudes.

Under the assumption that the whole stars in the G2 group have the same metallicity, which is reasonable because they have comparable \hkjwl\ strengths at a given $V$ magnitude, the bright RGBB magnitudes in the G2 \cni\ and \cns\ can be interpreted that they are enhanced in helium by \dy\ $\approx$ 0.03 -- 0.07 ($\pm$0.02)\footnote{We also derived a relation between RGBB magnitude and helium abundance using the results by \citet{valcarce12}, finding $\Delta m_{\rm bol} \approx  2.5\times\Delta Y$ for $Z = 1.6\times10^{-3}$ \citep[see][]{lee15}.} with respect to the G2 \cnw, which is marginally in agreement with the population synthesis model of M22 by \citet{joo13}, who suggested a helium enhancement of \dy\ = 0.09. It also should be mentioned that the fraction of the helium-enhanced population to explain the extreme blue horizontal branch (EBHB) population of M22 by \citet{joo13} was about 0.30, which is in good agreement with that of our helium-enhanced populations (i.e., the G2 \cni\ and \cns\ RGB stars, which eventually evolve into the EBHB phase), 0.28 $\pm$ 0.04.

\section{Summary and Conclusion}
Our \pcnjwl\ versus \pchjwl\ of M22 RGB stars shows discrete double CN--CH anticorrelations, which are due to metallicity difference between the two groups of stars as we already discussed in our previous work \citep{lee15}. Our populational number ratio of \nrgb\ = 63:37 ($\pm$3) is in excellent agreement with that by \citet{milone17}. 

The $\Delta_{\rm 2}$ distribution of the G1 (i.e., the lower-metallicity group) can be fitted best with a two-component model without a helium enhancement between the two subpopulations, namely, the G1 \cnw\ and \cns, inferred from their RGBB magnitudes, which is in sharp contrast to normal GCs with significant helium enhancements between the \cnw\ and \cns\ populations \citep[e.g., see][]{lee17,lee18,lee19b,largioia18,milone18}. On the other hand, the $\Delta_{\rm 2}$ distribution of the G2 (i.e., the higher-metallicity group) can be fitted best with three subpopulations, namely, the G2 \cnw, \cni, and \cns. The G2 appears to be more metal rich than the G1 by $\Delta$[Fe/H] $\approx$ 0.26 $\pm$ 0.05 dex. Unlike the G1 group, the G2 \cni\ and \cns\ appear to be enhanced in helium by \dy\ $\approx$ 0.03 -- 0.07 ($\pm$0.02) with respect to the G2 \cnw, a generic feature of normal GCs. The fraction of the G2 \cni\ and \cns, which are helium-enhanced subpopulations and will eventually evolve into the EBHB, is 0.28 $\pm$ 0.04, in good agreement with that estimated by \citet{joo13}.

The proper motion study from the \gaia\ DR2 allows us to reveal the the kinematical differences, in the sense that the G2 appears to rotate faster than the G1, confirming our previous results from the radial velocity measurements \citep{lee15}.

In both main groups, the CRDs of the \cns\ subpopulations are more centrally concentrated than other subpopulations. Interestingly, the CRDs of the G1 \cns\ and the G2 \cns\ are very similar. Likewise, the G1 \cnw\ and the G2 \cnw\ and \cni\ have almost identical CRDs.

In our previous study \citep{lee15}, we suggested that M22 most likely formed via a merger of two GCs,\footnote{Recently, \citet{massari19} suggested that M22 is an in situ GC.} based on the chemical, kinematical, and structural differences between the \caw\ (i.e., the G1 group of this study) and \cas\ (i.e. the G2 group) populations. It is believed that our results presented in this work also strongly support the idea of the merger scenario for M22. For example, the sequential formation scenario (e.g., in a formation sequence of G1 \cnw\ $\rightarrow$ G1 \cns\ $\rightarrow$ G2 \cnw\ $\rightarrow$ G2 \cni\ $\rightarrow$ G2 \cns, in which the metallicity evolution from the G1 to G2 groups and the helium enhancements in the G2 \cni\ and \cns\ can be explained, or in different sequences) cannot be reconciled with the kinematical properties and the CRDs of the subpopulations in the G1 and G2 groups. If so, the synchronization of the CRDs of the individual subpopulations between the G1 and G2 may hint that the CRDs of individual subpopulations in the G1 and G2 are mass-independent, but, perhaps, they are governed by some global processes. Future theoretical simulations based on new results of chemical, structural, and kinematical differences between MSPs will help to reveal the true story of M22.

\acknowledgements
J.-W.L.\ acknowledges financial support from the Basic Science Research Programs (grant Nos.\ 2016-R1A2B4014741 and NRF-2019R1A2C2086290) through the National Research Foundation of Korea (NRF). He also thanks Donghoh Kim for a helpful discussion and the anonymous referee for a constructive review of the manuscript.

\end{document}